# High-field magnetoelectric coupling and successive magnetic transitions in Mn-doped polar antiferromagnet Ni$_3$TeO$_6$

J. H. Zhang[1], L. Lin[1,2,*], C. Dong[3], Y. T. Chang[3], J. F. Wang[3], C. L. Lu[3], P. Z. Chen[1], W. J. Zhai[1], G. Z. Zhou[1], L. Huang[1], Y. S. Tang[4], S. H. Zheng[5], M. F. Liu[5], X. H. Zhou[1], Z. B. Yan[1], and J. -M. Liu[1]

[1]*Laboratory of Solid State Microstructures, Nanjing University, Nanjing 210093, China*
[2]*Department of Applied Physics, College of Science, Nanjing Forestry University, Nanjing 210037, China*
[3]*Wuhan National High Magnetic Field Center and School of Physics, Huazhong University of Science and Technology, Wuhan 430074, China*
[4]*School of Science, Nanjing University of Posts and Telecommunications, Nanjing 210023, China*
[5]*Institute for Advanced Materials, Hubei Normal University, Huangshi 435002, China*

* Corresponding author: llin@njfu.edu.cn

**[Abstract]** Among the 3$d$ transition metal ions doped polar $Ni_3TeO_6$, Mn-doped $Ni_3TeO_6$ has stimulated great interest due to its high magnetic ordering temperature and complex magnetic phases, but the mechanism of magnetoelectric (ME) coupling is far from understood. Herein we report our systematic investigation of the chemical control of magnetism, metamagnetic transition, and ME properties of $Ni_{3-x}Mn_xTeO_6$ single crystals in high magnetic field ($H$) up to 52 T. We present a previously unreported weak ferromagnetic behavior appeared in the $ab$ plane below 9.5 K in addition to the incommensurate helical and commensurate collinear antiferromagnetic states. In the low-field region, a spin-flop type metamagnetic transition without any hysteresis occurs at $H_{c1}$ for $H // c$, while another metamagnetic transition accompanied with a change in electric polarization is observed at $H_{c2}$ in the high-field region both for $H // c$ and $H // ab$ above 30 K, which can be attributed to the sudden rotation of magnetic moments at Ni2 sites. The ME measurements reveal that a first-order ME effect is observed in the low-$T$ and low-$H$ regions, while a second-order ME coupling term appears above 30 K in the magnetic field range of $H_{c1} < H < H_{c2}$ for $H // c$ and $H < H_{c2}$ for $H // ab$, both becoming significant with increasing temperature. Eventually, they are dominated by the second-order ME effect near the antiferromagnetic transition temperature. The present work demonstrates that $Ni_{3-x}Mn_xTeO_6$ is an exotic magnetoelectric material compared with $Ni_3TeO_6$ and its derivatives, thereby providing insights to better understand the magnetism and ME coupling in $Ni_3TeO_6$ and its derivatives.

**I. Introduction**

Multiferroics, in which more than one ferroic order coexist and couple with each other, have been extensively studied owing to their intriguing physics and great potential applications in multifunctional devices [1-7]. In particular, the strong cross-coupling between magnetic and ferroelectric orders gives rise to spin-driven ferroelectricity, large magnetoelectric (ME) coupling effect, and electrical manipulation of magnetic properties [8]. However, due to the strict constraints of crystal and magnetic symmetry [9,10], very few materials demonstrate strong ME coupling. Until recently, the polar magnets have been recognized as the promising playground to search for new multiferroics with large ME effects and emergent physical phenomena, such as giant thermal Hall effect [11], antiferromagnetic solutions [12,13], nonreciprocal effect [14,15], and topological quasiparticle skyrmions [16-19]. In contrast to the type-I multiferroics, e.g., $BiFeO_3$ [20] and hexagonal $YMnO_3$ [21], where the polar distortion is caused by nonmagnetic ions, the polar distortion in the polar ME materials is usually associated with the magnetic ions [22], thus strong magnetoelectric coupling can be expected.

Among the polar magnet such as $M_2Mo_3O_8$ ($M$ = 3$d$ transition metal) [23-25] and $CaBaCo_4O_7$ [26,27], the corundum-related compound $Ni_3TeO_6$ has attracted considerable interest due to its non-hysteresis colossal ME effect and successive metamagnetic phase transitions revealed by a high magnetic field [28]. It crystallizes in a noncentrosymmetric $R3$ space group with three distinct Ni crystallographic sites, denoted as Ni1, Ni2, and Ni3, respectively, as illustrated by VESTA software [29] shown in Fig. 1 (a). A collinear antiferromagnetic (AFM) order is established below Néel temperature $T_N$ ~ 52 K, accompanied with sudden changes of polarization at the magnetic transition, and a colossal magnetically induced electric polarization as large as 3280 $\mu C/m^2$ was obtained at 2 K [30]. Recent works have demonstrated that partially substituted $Ni^{2+}$ with other 3$d$ transition metal ions, such as $Mn^{2+}$ and $Co^{2+}$, can preserve the non-centrosymmetric space group $R3$, while the magnetic structure and ME coupling can be flexibly tuned [31-33]. This is probably attributed to the modified multiple exchange interactions illustrated in Fig. 1(a). For example, upon different $Co^{2+}$ doping, the magnetic structure can be modulated into an incommensurate helical phase with the spins lying in the $ab$ plane, allowing for largely tuned ME coupling [34]. A metamagnetic transition occurred at 7.5 T and 2 T for $Ni_2CoTeO_6$ and $NiCo_2TeO_6$, respectively, above which a linear ME effect with the ME coefficient of 3.4 ps/m and 323 ps/m was observed when magnetic field ($H$) was applied along the $ab$ plane. The magnetically induced electric polarization of 550 $\mu C/m^2$ for $Ni_2CoTeO_6$ and 5200 $\mu C/m^2$ for NiCo$_2$TeO$_6$ was obtained at 2 K, respectively.

In contrast to the Co$^{2+}$ doping studies, an incommensurate (IC) helical phase with spins lying in *ab* plane emerges between the AFM order temperature $T_N \sim 75$ K and $\sim 65$ K in powder sample Ni$_{2.1}$Mn$_{0.9}$TeO$_6$, below which a commensurate collinear (CC) antiferromagnetic phase with spins lying along the *c*-axis is established till to the lowest temperature [32]. In Figs. 1(b) and 1(c), we show the schematic magnetic structures of Ni$_{2.1}$Mn$_{0.9}$TeO$_6$ above and below 65 K, respectively. Compared with the significant easy-plane magnetic anisotropy observed in (Ni,Co)$_3$TeO$_6$, due to the small easy-axis anisotropy of Ni$^{2+}$ ions in Ni$_3$TeO$_6$ [28,35], the magnetic anisotropy of Ni$_{3-x}$Mn$_x$TeO$_6$ is reduced even further by substituting the Ni$^{2+}$ ions with isotropic Mn$^{2+}$. In consequence, the thermal fluctuations play a crucial role in the emergence of intermediate IC helical phase in addition to the exchange interaction and magnetic anisotropy [32]. It is noted that although the ME coupling and magnetism of Ni$_{3-x}$Mn$_x$TeO$_6$ have been recently reported by Kim *et al*, only one doping component (Ni$_{2.4}$Mn$_{0.6}$TeO$_6$) is provided, which is insufficient to explore the effect of site-specific doping on emergent magnetic phases and ME coupling effect. Therefore, two important questions are how the electric polarization evolves upon Mn doping, and whether are there any hidden exotic magnetic states by introducing chemical substitutions or external magnetic fields, although the microscopic mechanism of the ME response is far from understood yet. To clarify these issues, it is highly desirable to revisit the Ni$_{3-x}$Mn$_x$TeO$_6$ compounds by extending the Mn contents to uncover the intrinsic ME response and the underlying physics mechanism.

Motivated by the discussions mentioned above, we report the investigation of the magnetism and ME responses of Ni$_{2.4}$Mn$_{0.6}$TeO$_6$ and Ni$_2$MnTeO$_6$ single crystals in a high magnetic field up to $\sim 52$ T. In contrast to Ni$_3$TeO$_6$ and its derivates, we present a weak ferromagnetic behavior appeared below 9.5 K for *H // ab*. The ME response exhibits a strong dependence on temperature and the strength of the magnetic field. It is revealed that while the first-order ME effect governs at low-*T* and low-*H* region, the second-order term governs at high-*T* region. We discuss our experimental results based on the Heisenberg exchange-striction driven magnetoelastic coupling, therefore interpreting the origin of the ME coupling in Ni$_{3-x}$Mn$_x$TeO$_6$. Our results demonstrate the unique magnetic phases and ME effects of Ni$_{3-x}$Mn$_x$TeO$_6$, thereby providing insights into the strategy for the design of materials with large ME effects in the Ni$_3$TeO$_6$-type derivatives and other polar magnets.

**II. Experimental Details**

The single crystals we initially wanted to grow were $Ni_2MnTeO_6$ and $NiMn_2TeO_6$. Therefore, polycrystalline $Ni_2MnTeO_6$ and $NiMn_2TeO_6$ were prepared by the solid-state reaction method as described in previous reports [31]. In detail, stoichiometric amounts of high-purity NiO, $TeO_2$, and MnO were ground and sintered in a tubular furnace under the oxygen flow at 800 °C with intermediate grindings to ensure a complete reaction. Then the as-prepared polycrystalline $Ni_2MnTeO_6$ and $NiMn_2TeO_6$ were submitted to grow single crystals by the chemical vapor transport (CVT) method at 830 °C in the change zone and 750 °C in the growth zone for two weeks, using $TeCl_4$ as a transport agent. The as-grown single crystals are naturally hexagonal in geometry, with ~ 2 mm in diameter and 0.1 mm in thickness. Nevertheless, for CVT growth of single crystals, it is known that the product would be more or less deviated from the original source in chemical composition. Using x-ray energy dispersive spectroscopy (EDS) measurements, the crystals grown from $Ni_2MnTeO_6$ and $NiMn_2TeO_6$ polycrystalline powder were determined to be $Ni_{2.4}Mn_{0.6}TeO_6$ and $Ni_2MnTeO_6$, respectively, which will be discussed in detail later. The crystallinity was checked by the x-ray diffraction (XRD) using the Bruker D8 Advance x-ray diffractometer with Cu $K_\alpha$ radiation (wavelength $\lambda$ = 1.5406 Å) on the crushed crystals. The Rietveld refinement was performed using the GSAS program [36].

Subsequently, the well-prepared crystals were submitted for a series of characterizations. The temperature ($T$) dependence of the magnetic susceptibility along ($\chi_c$) and perpendicular ($\chi_{ab}$) to the $c$ axis was measured by the Quantum Design Superconducting Quantum Interference Device magnetometer (SQUID) in the zero-field-cooled (ZFC) and field-cooled (FC) modes with a measuring magnetic field $H$ = 0.1 T. The $H$-dependent magnetization ($M$) was measured at selected $T$ using Vibrating Sample Magnetometer (VSM) in Physical Property Measurement System (PPMS, Quantum Design). The specific heat from 5 K to 80 K was measured by a thermal-relaxation method using PPMS.

For the electrical and ME measurements [37], the crystals aligned with large surface normal to [001] were deposited with Au electrodes. The sample was poled in the electric field $E_{pole}$ = 20 kV/cm and selected $H$ from 120 to 10 K. Then the poling electric field was removed, followed by a sufficiently long short-circuit to avoid the influence of the injected charge. The $T$-dependent pyroelectric current ($J_{py}$) along the $c$ axis was measured using a Keithley 6514 electrometer with a temperature warming rate of 3 K/min. The electric polarization along the $c$ axis ($P_c$) was obtained by performing the integration of pyroelectric current with respect to time. Prior to the collection of magnetocurrent, the crystal was prepared under the ME poling procedure consisting of poling electric field $E_{pole}$ = 20 kV/cm and $H$ = 9 T with $H // c$ (or $H // ab$) during the cooling sequence down to selected temperature. Then the magnetocurrent was measured under selected $T$ upon $H$-ramping from 0→9 T→0 at a rate of 100 Oe/s.

For the high-field measurements, magnetization was measured using a coaxial pick-up coil while the magnetoelectric current was detected by a shunt resistor (10 kΩ) with a bias field of $E$ = 20 kV/cm applied. To ensure the accuracy of data, the baseline was measured under the same discharge voltage and the signals from the sample were calibrated by comparison with the low-field data measured by PPMS. In the high-field electric polarization measurements, the magnetoelectric current in the paramagnetic phase, i.e., $T$ = 80 K, was used as the baseline. All the high-field $M$ and magnetically induced polarization along $c$-axis ($\Delta P_c$) were measured using a 10.5 ms short-pulse magnet at Wuhan National High Magnetic Field Center (WHMFC).

**III. Results and Discussion**

*A. Single crystal characterization*

Before presenting the magnetic data, the elemental composition of the as-grown single crystals was determined carefully. The EDS spectroscopy of single crystals grown from $Ni_2MnTeO_6$ and $NiMn_2TeO_6$ polycrystalline powder, as shown in Figs. 2(a) and 2(b), gives rise to the atomic Ni : Mn : Te : O ratio of 2.36 : 0.64 : 1.01 : 6.86  and 1.96 : 1 : 1 : 6.56, respectively, which is very close to the stoichiometry of $Ni_{2.4}Mn_{0.6}TeO_6$ and $Ni_2MnTeO_6$. Details of the EDS mapping images and spectrums results are included in the Supplemental Materials [38]. In addition, the chemical composition obtained from the Rietveld refinement of crushed $Ni_{2.4}Mn_{0.6}TeO_6$ and $Ni_2MnTeO_6$ single crystal is determined to be $Ni_{2.3}Mn_{0.7}TeO_6$ and $Ni_2MnTeO_6$, respectively, in agreement with the elemental composition determined using EDS. Figure. 2(c) present the room-temperature x-ray diffraction patterns onto the large hexagonal plane of the as-grown $Ni_{2.4}Mn_{0.6}TeO_6$ and $Ni_2MnTeO_6$ crystals. The well-defined (00L) reflections, as well as the full width at half-height (FWHH) of 0.24º for the (006) peak, indicate good crystallinity and chemical homogeneity of our samples. In Fig. 2(d) we show the slow-scan x-ray diffraction spectra and Rietveld refinement of crushed $Ni_2MnTeO_6$ single crystals. The refined structure of $Ni_2MnTeO_6$ fits the non-centrosymmetric trigonal ($R3$) symmetry with the unit cell parameters of $a = b = 5.1727(8)$ Å and $c = 14.0196(4)$ Å. More detailed structural parameters of $Ni_{2.4}Mn_{0.6}TeO_6$ and $Ni_2MnTeO_6$ determined from the Rietveld refinement of XRD data of crushed crystals are summarized in Table I. It is revealed that the Ni1 and Ni3 crystallographic sites are occupied by the Ni ions while the Ni2 sites are mainly occupied by the Mn ions, in agreement with previously reported data from powder neutron scattering [31,32].

*B. Magnetic susceptibility and specific heat*

The $T$ dependence of magnetic susceptibility ($\chi = M/H$) parallel ($\chi_c$) and perpendicular ($\chi_{ab}$) to the $c$-axis under $H = 0.1$ T for $Ni_{2.4}Mn_{0.6}TeO_6$ and $Ni_2MnTeO_6$ single crystals are shown in Figs. 3(a) and 3(b), respectively. For $H // c$, we find a long-range antiferromagnetic ordering established at $T_N = 72.5$ K for $Ni_{2.4}Mn_{0.6}TeO_6$, while this value is slightly enhanced to $T_N = 77$ K for $Ni_2MnTeO_6$, the highest one among the pure and doped $Ni_3TeO_6$ series to the best of our knowledge. It is evident that in $Ni_{3-x}Mn_xTeO_6$ larger content of Mn doping improves the AFM transition temperature, while unfortunately the second impurity phase $Mn_3TeO_6$ emerges when $x > 1$, and increases in intensity with increasing $x$. Hence, the crystal $Ni_2MnTeO_6$ is the ideal candidate to be studied in terms of magnetic property and ME coupling effects. Upon cooling from $T_N$, another kink shows up at $T_1 = 65$ K, accompanied by a sharp drop between $T_1 = 65$ K and $T_2 = 62$ K for $Ni_{2.4}Mn_{0.6}TeO_6$. Correspondingly, the $\chi_{ab}$ shows an unambiguous hump at $T_N$ and upward steps between $T_1$ and $T_2$. Furthermore, we plot the d$\chi_{ab}$/d$T$ curve in the inset of Fig. 3(b) to better illustrate these transition points. Interestingly, a clear ferromagnetic enhancement in $\chi_{ab}$ is observed below $T_3 = 9.5$ K. The value of $M_{ab}$ at 2 K in $H = 0.1$ T is 0.064 $\mu_B$/f.u. for $Ni_{2.4}Mn_{0.6}TeO_6$ and 0.144 $\mu_B$/f.u. for $Ni_2MnTeO_6$, corresponding to 1.64% and 3.33% of the perfectly saturated values of the magnetic system. The small value of $M_{ab}$ immediately eliminates the possibility of normal ferromagnetic long-range order. Since the sample is highly insulating, itinerant or band ferromagnetism also is impossible. Based on the consideration mentioned above, due to the decreased easy-axis magnetic anisotropy, the ferromagnetic enhancement in $\chi_{ab}$ should be caused by the small canted magnetic moments in $ab$ plane when the measuring magnetic field of 0.1 T is applied along the *ab* plane.

According to the neutron diffraction measurements on powder sample $Ni_{2.1}Mn_{0.9}TeO_6$ [32], the first anomaly at $T_N$ in $\chi(T)$ curves corresponds to the incommensurate spin-helical order, where the spins lie in the *ab* plane and stack along the *c* axis, while the second one is attributed to the spin-reorientation (SR) transition. This suggests that the alignment of magnetic moments rotates from the *ab* plane at $T_1$ and finally aligns to the *c* axis at $T_2$, forming a commensurate collinear (CC) antiferromagnetic order along the *c* axis below $T_2$. In fact, the temperature $T_1$ and $T_2$ for the SR transition is not a fixed value, and it changes depending on the direction and the magnitude of *H*. As shown in Fig. 3(c), for *H* // *c*, with increasing *H*, the temperature $T_1$ and $T_2$ for $Ni_{2.4}Mn_{0.6}TeO_6$ shift to the lower *T* and disappear above 6 T while the temperature $T_N$ is almost unchanged. On the contrary, for *H* // *ab*, the temperature $T_1$ and $T_2$ shift to the higher *T* and emerge with $T_N$ above 4 T as demonstrated in the Fig. 3(d). Here, regarding the nature of *H*-dependent temperatures of SR transition, we speculate that the Landau free analysis is necessary to explain this phenomenon [39], and additional experiments are currently underway to ensure the reliability of the analysis.

To further investigate this interesting *H*-modulated magnetic property, the specific heat ($C_p/T$) of $Ni_{2.4}Mn_{0.6}TeO_6$ was measured under selected *H* = 0.1 and 4 T applied along the *c* axis, as shown in Figs. 3(e) and 3(f) where the $\chi(T)$ data with the same magnitude of *H* applied along the *c* axis are plotted as well for reference. The sharp $\lambda$-type peak in $C_p/T$ curves was observed at $T_N$ = 72.5 K, indicating that the first magnetic transition is a second-order transition. In addition, another small change was also observed at $T_1$ = 65 K for *H* = 0.1 T and $T_1$ = 45 K for *H* = 4 T, in accordance with the anomaly in the $\chi(T)$ curves.

*C. Low-field magnetization*

In Fig. 4(a) we first show the *H*-dependent magnetization of $Ni_{2.4}Mn_{0.6}TeO_6$ up to 9 T in the *H* // *c* geometry at various temperatures. A clear magnetic field-induced metamagnetic transition is identified at critical magnetic field $H_{c1}$ (marked by black arrows), characterized by the abrupt increase of $M_c(H)$ without any hysteresis, quite similar to the parent phase $Ni_3TeO_6$. It is worth noting that in $Ni_3TeO_6$ such a metamagnetic transition was attributed to the spin-flop transition and was designated as a second-order transition [30]. However, the single crystal neutron diffraction shows that this metamagnetic transition is in fact a first-order transition from a commensurate collinear antiferromagnetic structure with spins along the $c$ axis to incommensurate conical spiral structure with a significant spin component lying in the $ab$ plane [35]. Therefore, considering the similarity with the $Ni_3TeO_6$, the magnetic structure of $Ni_{3-x}Mn_xTeO_6$ above $H_{c1}$ is supposed to be conical spiral structure with major spins lying in the $ab$ plane. As shown in Fig. 4(b), the values of $H_{c1}$, determined by $dM_c/dH$ for $Ni_{2.4}Mn_{0.6}TeO_6$ and $Ni_2MnTeO_6$, fall within the range of 2 ~ 4 T, dramatically lowered compared with that of 6 ~ 8 T in $Ni_3TeO_6$. The decreased value of $H_{c1}$ in $Ni_{3-x}Mn_xTeO_6$ should be caused by the decreased easy-axis magnetic anisotropy due to the isotropic $Mn^{2+}$ ions doping.

In contrast to the non-hysteresis metamagnetic transition observed for $H // c$, the magnetization in the $H // ab$ ($M_{ab}$) geometry exhibits linear increase with $H$ above $T \sim 10$ K, while at low temperature as shown in Fig. 4(c), a characteristic steep increase of $M_{ab}$ is observed, which has never been reported in previous $Ni_3TeO_6$ and Co-doped crystals. To explore more details of the phenomenon, we present the $H$-dependent magnetization of $Ni_{2.4}Mn_{0.6}TeO_6$ and $Ni_2MnTeO_6$ at $T = 5$ K enlarged to the low field region $H < 2$ T in Fig. 4(d). For $H // c$, $M_c$ increases linearly with a magnitude much smaller than that of $M_{ab}$, further indicating the presence of small canted magnetic moments in the $ab$ plane under the small magnetic field applied along the $ab$ plane.

*D. Magnetoelectric coupling in the low field*

Figures. 5(a) and 5(b) display the $T$-dependent pyroelectric current density $J_{py}(T)$ along the $c$ axis for $Ni_{2.4}Mn_{0.6}TeO_6$ under different $H$ applied along and perpendicular to the $c$ axis, respectively. Consistent with the two anomalies observed in $\chi(T)$ curves shown in Figs. 3(c) and 3(d), two sharp pyroelectric current peaks appear at $T_1 \sim 69$ K and $T_2 \sim 63$ K for $H = 0$. As the magnetic field increases, for $H // c$, the lower-$T$ peak near $T_2$ shifts to lower-$T$ side and disappears above 6 T, while the higher-$T$ peak at $T_1 \sim 69$ K is almost unchanged. On the contrary, for $H // ab$, the lower-$T$ peak gradually moves towards the higher temperature direction and eventually fuses with $T_1$ as the magnetic field increases. The same evolution of the current peak in $J_{py}(T)$ and anomalies in $\chi(T)$ response to $H$ indicate that the higher-$T$ peak in $J_{py}(T)$ is induced by AFM ordering, while the lower-$T$ peak comes from SR transition. The change in the electric polarization $\Delta P_c$ at 10 K for $H = 0$ is increased from 1152 $\mu C/m^2$ for $Ni_{2.4}Mn_{0.6}TeO_6$ to 1950 $\mu C/m^2$ for $Ni_2MnTeO_6$ as shown in Figs. 5(c) and 5(e), indicating that both magnetic ordering temperature and electric polarization can be enhanced upon increased Mn-doping level. It is worth noting that the orientation of the magnetic field has almost no effect on the intensity of $c$ axis ferroelectric polarization as shown in Figs. 5(c) to 5(f).

To further demonstrate the ME response of $Ni_{2.4}Mn_{0.6}TeO_6$ and $Ni_2MnTeO_6$, the change of electric polarization $\Delta P_c = P_c(H) - P_c(H = 0)$ along the $c$ axis at various $T$ as a function of $H$ was measured under the $H$ applied along and perpendicular to the $c$ axis, respectively, shown in Figs. 6(a)-6(d). Consistent with the stepwise sudden increase in $M_c(H)$ at $H_{c1}$, the jump of $\Delta P_c$ at $H_{c1}$ can be clearly seen in the $H$ scanning, indicating the intrinsic magnetic origin of $\Delta P_c$. Firstly, as the doping level increases, the polarization intensity increases from 243 $\mu C/m^2$ for $Ni_{2.4}Mn_{0.6}TeO_6$ to 397 $\mu C/m^2$ for $Ni_2MnTeO_6$ at $T = 10$ K for $H // c$. For comparison, we have grown the parent phase $Ni_3TeO_6$ crystals using the same growth procedure, and the magnetically induced polarization $\Delta P_c$ reaches ~ 200 $\mu C/m^2$ at 10 K [34], implying that samples with higher Mn doping ratio are more advantageous in enhancing the magnetoelectric coupling response, e.g., twice the amplitude of the parent phase in $Ni_2MnTeO_6$. In contrast, as shown in Figs. 6(b) and (d), no sharp jump of $\Delta P_c$ is observed under $H // ab$ due to the absence of metamagnetic transition. Secondly, it shows that under $H // c$ the linear ME effect is dominated in the magnetic field range of $H_{c1} < H < 9$ T below 30 K, while with the $T$ increasing above 30 K, a second-order ME effect gradually emerges as illustrated by the fitting lines. A similar phenomenon is also observed for $\Delta P_c$ measured under $H // ab$.

In general, the electric polarization of a ME material is given as a function of $H$ up to the second order as

$$P = P_{latt} + P_{spin} + \alpha H + \beta H^2, \qquad (1)$$

where $P_{latt}$ and $P_{spin}$ are the lattice- and spin-induced spontaneous polarization, respectively, which are finite even in zero field. The coefficients $\alpha$ and $\beta$ represent the first- and second-order ME coefficients, respectively. Noting that no absolute electric polarization can be obtained from the magnetoelectric current measurements, and the obtained polarization is the change of electric polarization. Consequently, the change of the electric polarization $\Delta P$ induced by the magnetic field can be expressed as

$$\Delta P = P_0 + \alpha H + \beta H^2, \qquad (2)$$

where $P_0$ is the difference of $P(H)$ and $P(H = 0)$ in the same magnetic phase and usually is zero. When a magnetic field-induced metamagnetic transition occurs, $P_0$ is the difference of $P(H=0)$ between the two different magnetic phases [40].

To qualitatively characterize the ME coupling properties of $Ni_{3-x}Mn_xTeO_6$, the ME coefficients of $Ni_2MnTeO_6$ are evaluated as representative by fitting the $\Delta P_c(H // c)$ and $\Delta P_c(H // ab)$ data using Eq. (2), as depicted in Figs. 6(c) and 6(d), respectively. The $\Delta P_c(H // c)$ curves above $H_{c1}$ can be well fitted by using Eq. (2), and the ME coefficients $\alpha$ and $\beta$ are highly dependent on the temperature. For instance, at $T = 10$ K, fitting the $\Delta P_c(H // c)$ curve leads to $\alpha = 40.3$ ps/m and $\beta = 0.06 \times 10^{-18}$ s/A, in stark contrast to the coefficients $\alpha$ and $\beta$ of 0.065 ps/m and $4.7 \times 10^{-18}$ s/A, respectively at $T = 70$ K. More obtained ME coefficients $\alpha$ and $\beta$ as a function of $T$ are presented in Figs. 6(e)-(f) for $H // c$ and $H // ab$, respectively. These results clearly indicate that the first-order ME effect dominates at low-$T$ region, while the second-order ME effect appears in the high-$T$ region and gradually takes the dominant role with the $T$ increasing.

*E. Metamagnetic transition and ME effect in high field*

To further elucidate the magnetic and ME properties, subsequently we will focus on exploring possible magnetic phase transitions and ME coupling in pulsed high magnetic fields, taking $Ni_2MnTeO_6$ again as an example. Figures 7(a) and 7(b) show the $H$ dependence of magnetization $M$ under the $H$ applied along ($M_c$) and perpendicular ($M_{ab}$) to the $c$ axis up to 52 T at selected temperatures, respectively. At $T = 10$ K, a clear stepwise increase in $M_c$ is evident at $H_{c1} = 3.9$ T, consistent with that observed from the low field magnetic measurement. Then, $M_c$ increases linearly with $H$ up to 25 T, noting that the linear extrapolation of the $M_c(H)$ data between 4 and 25 T gives a finite intercept of 0.04 $\mu_B$/f.u. at $H = 0$, which is consistent with the emergence of the helical conical magnetic structure mentioned earlier. As the $H$ further increases, the slope of the $M_c(H)$ curve decreases slightly. The value of $M_c$ at 57 T is 4.75 $\mu_B$/f.u., which is much smaller than the expected saturation magnetization $M_S = 9$ $\mu_B$/f.u. for two $Ni^{2+}$ ions with $S = 1$ and one $Mn^{2+}$ ion with $S = 5/2$ (assuming gyromagnetic ratio $g = 2$). Interestingly, when the temperature increases up to 30 K, evidenced by the anomaly (marked by black arrows) in the derivative $dM_c/dH$ curve at 30 K in Fig. 7(c), anomaly assigned as another metamagnetic transition is observed at $H_{c2} = 47$ T. The value of $H_{c2}$ for the metamagnetic transition decreases monotonically with the $T$ increasing, finally disappears at 75 K (not shown). For $H // ab$, a similar metamagnetic transition is observed at $H_{c2}$ above 30 K, shown in Figs. 7(b) and 7(d).

In Figs. 8(a) and 8(b), we present the $H$ dependence of $\Delta P_c$ along the $c$ axis under $H // c$ and $H // ab$ cases at selected $T$, respectively. For $H // c$, the low-field region of $\Delta P_c(H)$ curves well reproduce the results in Figs. 6(c). Concomitant with the $H$-induced metamagnetic transition at high field, $\Delta P_c$ exhibits a significant change at $H_{c2}$, evidenced by the broad peak (marked by black arrows) in the $H$ dependent magnetoelectric current shown in Fig. S5 in the Supplementary Material [38]. It is noteworthy that the hysteretic change of $\Delta P_c$ at $H_{c2}$ may be correlated with the magnetoelectric current measured at various temperatures, utilizing the same baseline measured at 80 K. In addition, similar colossal magnetoelectricity is also observed for $H // ab$. Overall, the magnitude of $\Delta P_c$ reaches ~ 0.31 μC/cm$^2$ at $T$ = 30 K and 52 T for $H // c$ and $H // ab$, larger than most ME materials. Besides, the one-to-one correspondences between $\Delta P_c(H)$ and $M_c(H)$ curves strongly confirms the intrinsic coupling between the electric polarization and magnetism in Ni$_2$MnTeO$_6$.

To provide a detailed quantitative characterization of the ME coupling properties of Ni$_2$MnTeO$_6$ in high field, the $\Delta P_c(H // c)$ and $\Delta P_c(H // ab)$ curves at $T$ = 10 and 30 K in the field-down run as representatives are fitted by the Eq. (2), shown in Fig. 8(c). For the case of $T$ = 10 K, the $\Delta P_c(H // c)$ curve can be well fitted within the magnetic field range of 3.9 T < $H$ < 15 T and 20 T < $H$ < 52 T, with the fitting parameters of $\alpha$ = 36.6 ps/m, $\beta$ = 0.18 ×10$^{-18}$ s/A, and $\alpha$ = 40.1 ps/m, $\beta$ = 1.08 × 10$^{-18}$ s/A, respectively. This indicates that in the low-$T$ region, the ME effect in low-$H$ region is primarily governed by the first-order ME effect, whereas in the high-$H$ region, a combination of the first- and second-order ME effects are governed. For $T$ = 30 K, both $\Delta P_c(H // c)$ and $\Delta P_c(H // ab)$ curves can be well fitted in the range of $H_{c1}$ < $H$ < $H_{c2}$ with $\alpha$ = 31.8 ps/m, $\beta$ = 0.99 × 10$^{-18}$ s/A and $\alpha$ = 19.7 ps/m, $\beta$ = 0.87 × 10$^{-18}$ s/A, respectively. More detailed ME coefficients $\alpha$ and $\beta$ as a function of temperature for $H // c$ are shown in Fig. 8(d). Consistent with the results observed in our steady-field measurements, the value of $\alpha$ decreases while $\beta$ increases with $T$ increasing, indicating the second-order ME effect becomes more significant with the $T$ increasing. Similar phenomenon is also observed in the $H // ab$ case. Moreover, it is worth noting that the slight difference in the ME coefficient between the steady field and high field may be attributed to the utilization of different methods.

*F. Discussion*

As mentioned above, the ME response exhibits a strong dependence on temperature and the strength of the magnetic field in $Ni_{3-x}Mn_xTeO_6$. Now, we firstly revisit the ME coupling mechanism in $Ni_3TeO_6$. $Ni_3TeO_6$ crystallizes in a polar $R3$ space group with three magnetic $Ni^{2+}$ ions and one nonmagnetic $Te^{6+}$ ion arranged along the $c$-axis. The Ni and Te ions, surrounded by six oxygens, form two face-sharing Ni3O6-TeO6 and Ni2O6-Ni1O6 octahedra dimers, separated by the octahedra vacancy. The electrostatic repulsion between the cations within the face-sharing octahedra dimers, will push them farther away from each other and closer to the vacancy, leading to the formation of dipole along the $c$-axis in each octahedra dimer and macroscopic polarization along the $c$-axis. In the magnetic order state, the Heisenberg exchange interactions between the spins will induce the magnetic ions to shift [41]. The shifts of charged magnetic ions in a polar structure results in the distortion of octahedra in the $ab$ plane, namely magnetoelastic coupling. This phenomenon leads to changes of electric polarization compared to the paramagnetic state. The changes in the electric polarization $\Delta P_c$ along the $c$ axis, can be expressed by Eq. (3), assuming that the Heisenberg exchange-striction dominates [28].

$$\Delta P_c = \sum_n \alpha_n S_n \cdot S_{n'}, \tag{3}$$

where $S_n$ and $S_{n'}$ are the spins connected by the exchange interaction $J_n$ shown in Fig. 1(a), $\alpha_n$ are the exchange-striction coefficient strongly dependent on the bond angles and bond lengths of neighboring ions, which are different in AFM and conical spiral magnetic phase.

After clarifying the ME coupling mechanism in $Ni_3TeO_6$, we now turn our attention to discuss the ME response observed in $Ni_{3-x}Mn_xTeO_6$. Firstly, we discuss the step-like changes of $\Delta P_c$ observed at $H_{c1}$. During the spin-flop type phase transition around $H_{c1}$, the spins rotates rapidly from the $c$ axis to $ab$ plane as well as the changes in $\alpha_n$, leading to the sudden changes in electric polarization, according to Eq. (3).

Secondly, we will discuss the temperature dependent ME response. When the applied $H$ along the $c$ axis exceeds $H_{c1}$, the magnetic structure of $Ni_{3-x}Mn_xTeO_6$ will change into the conical spiral structure. In this phase, according to Eq. (3), the change of electric polarization $\Delta P_c$ can be expressed as

$$\Delta P_c = \sum_n \alpha_n S_n \cdot S_{n'} = \sum_n \alpha_n S_n^{ab} \cdot S_{n'}^{ab} \cos\theta_n + \sum_n \alpha_n S_n^z \cdot S_{n'}^z, \tag{4}$$

where $S_n^{ab}$ and $S_n^c$ represent the $ab$-plane and $z$-components of $S_n$, respectively, and $\theta_n$ denotes the angle between $S_n^{ab}$ and $S_{n'}^{ab}$ in $ab$-plane. As shown in Fig. 7(a), the magnetization parallel to the $c$ axis, i.e., $M_c = \sum_n S_n^z$, changes linearly with $H$ in low-$H$ region, implying that $S_n^z$ increases linearly with increasing $H$. Therefore, the $H^2$ term should appear in the expression of $\Delta P_c$. However, as the spins are mainly lying in the $ab$-plane with small $z$-components at low-$T$ and low-$H$ region in the conical spiral phase, the quadratic term can be ignored. Thus, the linear variation of $\Delta P_c$ with respect to $H$ should arise from alterations in the $ab$-plane components of the spins under the applied magnetic field. In contrast, the $z$-components of $S_n^c$ will increase dramatically under high magnetic field, and cannot be ignored anymore, leading to the gradual emergence and enhancement of the quadratic term $H^2$.

As the temperature increases, the spins can more easily tilt towards the $c$ axis under the influence of an applied magnetic field along the $c$ axis. As a result, the influence of the first-order ME effect weakens, while the significance of the second-order ME effect increases, ultimately dominating around $T_N$. In the case of $H // ab$, the spins $S_n$ will tilt slightly towards the $ab$-plane, causing a small $ab$-component of magnetic moments to emerge in the $ab$-plane. Since the magnetization parallel to the $ab$-plane is proportional to $H$ in the low-$H$ region, the $H^2$ term can also appear in the expression of $\Delta P_c$. Similar conclusions can also be drawn in this scenario.

Finally, we shift our focus to the physical interpretation of the second metamagnetic transition at $H_{c2}$. Given that the high-field metamagnetic transition observed in $Ni_3TeO_6$ is governed by the competition between Zeeman and exchange energies, it is imperative to identify the effect of Mn-doping on the exchange interactions and how it affects these interactions. By comparing the crystal structure parameters of $Ni_2MnTeO_6$ with that of $Ni_3TeO_6$, which were grown under the same conditions. It is evident that the Mn-doping has a more pronounced effect on the magnetic exchange interaction $J_2$ and $J_4$ as illustrated in Fig. 1(a). In detail, concerning the ferromagnetic exchange interaction $J_2$, although the bond angle remains almost unchanged, the bond length associated with $J_2$ in $Ni_2MnTeO_6$ is significantly elongated. This elongation could be attributed to the substitution of $Ni^{2+}$ ions at Ni2 sites by larger $Mn^{2+}$ ions. Hence, the exchange interaction $J_2$ should be weakened during the Mn doping. In contrast, regarding the antiferromagnetic exchange interaction $J_4$ (the strongest one in $Ni_3TeO_6$), the bond angle increases significantly from 131° in $Ni_3TeO_6$ to 137° in $Ni_2MnTeO_6$, while the bond lengths remain almost unchanged, implying that the exchange interaction $J_4$ is enhanced during Mn doping. Therefore, it is reasonably to speculate that the increased magnetic ordering temperature in $Ni_{3-x}Mn_xTeO_6$ should be caused by the strengthened AFM interaction $J_4$.

After identify the effect of the Mn-doping on the magnetic exchange interactions, we would like to discuss the possible origin of the high-field metamagnetic transition occurred at $H_{c2}$ in Ni$_2$MnTeO$_6$. Considering the resemblance of the $\Delta P_c(H)$ and $M(H)$ curves between Ni$_3$TeO$_6$ and Ni$_2$MnTeO$_6$, the metamagnetic transition occurred at $H_{c2}$ in Ni$_2$MnTeO$_6$ should correspond to the one occurred at 35 T in Ni$_3$TeO$_6$ [15,41,42], which is caused by the sudden rotation of the spins of Mn$^{2+}$ ions at Ni2 sites towards to applied $H$ [28]. In particular, considering the similarity between the (Ni,Mn)$_3$TeO$_6$ and Ni$_3$TeO$_6$, a stronger ME coupling in higher field can be expected. Unfortunately, a higher field is not accessible to us at this stage and deserves further investigation.

Besides the Heisenberg exchange-striction mechanism, other mechanisms such as ferro-axial [43] and spin-dependent $p$-$d$ hybridization mechanism [44], are also considered. In the paramagnetic phase, the crystal structure of Ni$_{3-x}$Mn$_x$TeO$_6$ belongs to the 3 ferroaxial class with a macroscopic axial vector A along the 3-fold axis [45], i.e., the $c$ axis. Similar to other ferroaxial multiferroic materials such as CaMn$_7$O$_{12}$ [43] and Cu$_3$Nb$_2$O$_8$ [46], in the conical spiral phase, the chirality of the conical spiral structure $\sigma = r_{nn'} \cdot (S_n^{ab} \times S_{n'}^{ab})$ couples to A and induces a polarization $P \propto \gamma \sigma A$ along the $c$ axis. Here, $S_n^{ab}$ and $S_{n'}^{ab}$ is $ab$-plane components of the spins $S_n$ and $S_{n'}$, respectively, $r_{nn'}$ is the bond vector connecting $S_n$ and $S_{n'}$, and $\gamma$ is the purely structural coupling constant. As the $H$ increases, there is a decrease in the $ab$-plane components of the spins, leading to a reduction in the contribution of the electric polarization from the ferro-axial mechanism.

In addition, according to Ref. [47], for the NiO$_6$ octahedral with threefold rotation symmetry, the contribution of the spin-dependent $p$-$d$ hybridization mechanism to $\Delta P_c$ can be expressed as:

$$\Delta P_c^{pd} = C_1 S^2 \sin^2\theta + C_2 S^2 \cos^2\theta = C_1 S^2 + (C_2 - C_1) S_z^2 \tag{5}$$

where $C_1$ and $C_2$ are the coupling constants correlated to the structure parameters of NiO$_6$ octahedral, $S$ is the magnetic moment at Ni sites, $\theta$ is the angle of spins with respect to the $z$-axis. From Eq. (5) we can obtain that only the $z$-component of the magnetic moment can affect $\Delta P_c$. Similarly, the $H^2$ term also appears in the expression of $\Delta P_c^{pd}$. At lower temperatures, as the majority of magmatic moments lie in the $ab$ plane, the contribution of $\Delta P_c^{pd}$ to the electric polarization can be ignored. As the $S_z$ increases, the contribution of $\Delta P_c^{pd}$ to the electric polarization also increases.

## IV. Conclusion

In conclusion, we have presented our systematic investigation on the magnetism and magnetoelectric coupling of $Ni_2MnTeO_6$ and $NiMn_2TeO_6$ single crystal in the field up to 52 T. In addition to the incommensurate helical and commensurate collinear antiferromagnetic states, below $T_3$ ~ 9.5 K a weak ferromagnetism is observed in the *ab* plane. The high-field study of $Ni_2MnTeO_6$ reveals another metamagnetic transition emerging above $T$ = 30 K. Such transition is caused by the sudden rotation of magnetic moments on the Ni2 site towards the magnetic field. In contrast to the parent phase $Ni_3TeO_6$, the ME effect is highly dependent on temperature and magnetic field. Our results unambiguously demonstrate that the first-order ME effect is observed in the low-$T$ and low-$H$ while the second-order ME effect is observed in the high-$T$ region and becomes significant with the $T$ increasing. We discuss our experimental results with the well-known mechanisms for spin-induced electric polarization, thereby clarifying different mechanisms of ME responses. Our results provide insights into the strategy for the design of materials with large ME effects in the $Ni_3TeO_6$-type derivatives and other polar magnets.

**Acknowledgment**

The authors would like to acknowledge the financial support from the National Natural Science Foundation of China (Grants No. 92163210, No. 12274231, No. 11834002, No. 12074135, No. 12304124, No. 12074111, No. 52272108, No. 51721001, and No. 11974167).

TABLE I. Structural parameters of $Ni_{2.4}Mn_{0.6}TeO_6$ (Mn0.6) and $Ni_2MnTeO_6$ (Mn1) determined from the Rietveld refinement of XRD data of crushed crystals at room temperature.

| sample | $a$ (Å) | $b$ (Å) | $c$ (Å) | $\alpha$ (°) | $\beta$ (°) | $\gamma$ (°) |
|---|---|---|---|---|---|---|
| Mn0.6 | 5.1438(3) | 5.1438(3) | 13.9101(6) | 90 | 90 | 120 |
| Mn1 | 5.1727(8) | 5.1727(8) | 14.0196(4) | 90 | 90 | 120 |
| | Atom | $x$ | $y$ | $z$ | Wyckoff | occupation |
| | Ni3/Mn3 | 1/3 | 2/3 | 0.1693(5) | 3a | 1/0 |
| | Ni2/Mn2 | 1.0000 | 1.0000 | 0.2140(7) | 3a | 0.44(2)/0.56(2) |
| Mn0.6 | Ni1/Mn1 | 2/3 | 1/3 | 0.3501(2) | 3a | 0.93(1)/0.07(3) |
| | Te | 1/3 | 2/3 | 0.3753(1) | 3a | 0.97(6) |
| | O1 | 0.6034(2) | 0.6627(5) | 0.2817(7) | 9b | 1.01(2) |
| | O2 | 0.9735(1) | 0.6712(7) | 0.4551(8) | 9b | 0.98(8) |
| | Ni3/Mn3 | 1/3 | 2/3 | 0.1631(2) | 3a | 0.93(2)/0.07(2) |
| | Ni2/Mn2 | 1.0000 | 1.0000 | 0.2135(4) | 3a | 0.18(2)/0.82(3) |
| | Ni1/Mn1 | 2/3 | 1/3 | 0.3398(6) | 3a | 0.90(1)/0.10(2) |
| Mn1 | Te | 1/3 | 2/3 | 0.3651(3) | 3a | 0.98(4) |
| | O1 | 0.6314(2) | 0.6650(0) | 0.2792(6) | 9b | 0.98(6) |
| | O2 | 0.9979(4) | 0.6418(5) | 0.4419(4) | 9b | 0.99(2) |
| | $R_p$ (%) | | $R_{wp}$ (%) | | $\chi^2$ | |
| Mn0.6 | 5.68 | | 3.78 | | 2.69 | |
| Mn1 | 4.95 | | 3.66 | | 1.82 | |

**Figures and Captions**

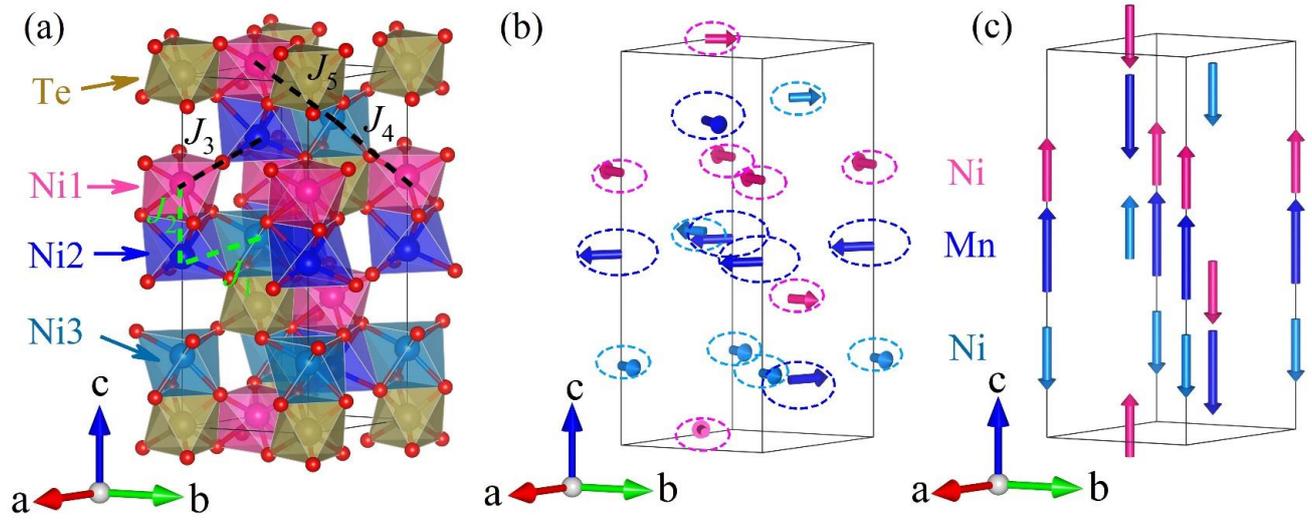

Fig. 1. (a) Crystal structure of $Ni_3TeO_6$ with Ni ions occupied three distinct crystallographic sites. The three Ni-oxygen (red spheres) cages are shown in pink, blue and light cyan, respectively. The Te-oxygen cages are shown in dark yellow. Here, green and black dashed lines denote the ferromagnetic ($J_1$ and $J_2$) and antiferromagnetic ($J_3$, $J_4$ and $J_5$) exchange interactions, respectively. The magnetic structure of (b) intermediate helical phase and (c) collinear states of $Ni_{2.1}Mn_{0.9}TeO_6$, drawn based on Ref. [32]. The pink, blue and light cyan arrows denote the spins on Ni1, Ni2, and Ni3 sites, respectively. The dashed ellipses in (b) illustrate the incommensurate helical spins propagating along the *c* axis. One structure unit cell is shown for simplicity. When moving from one cell to the next along the *c*-axis, the spins rotate by 120° in the *ab* plane (b) and by 180° in (c).

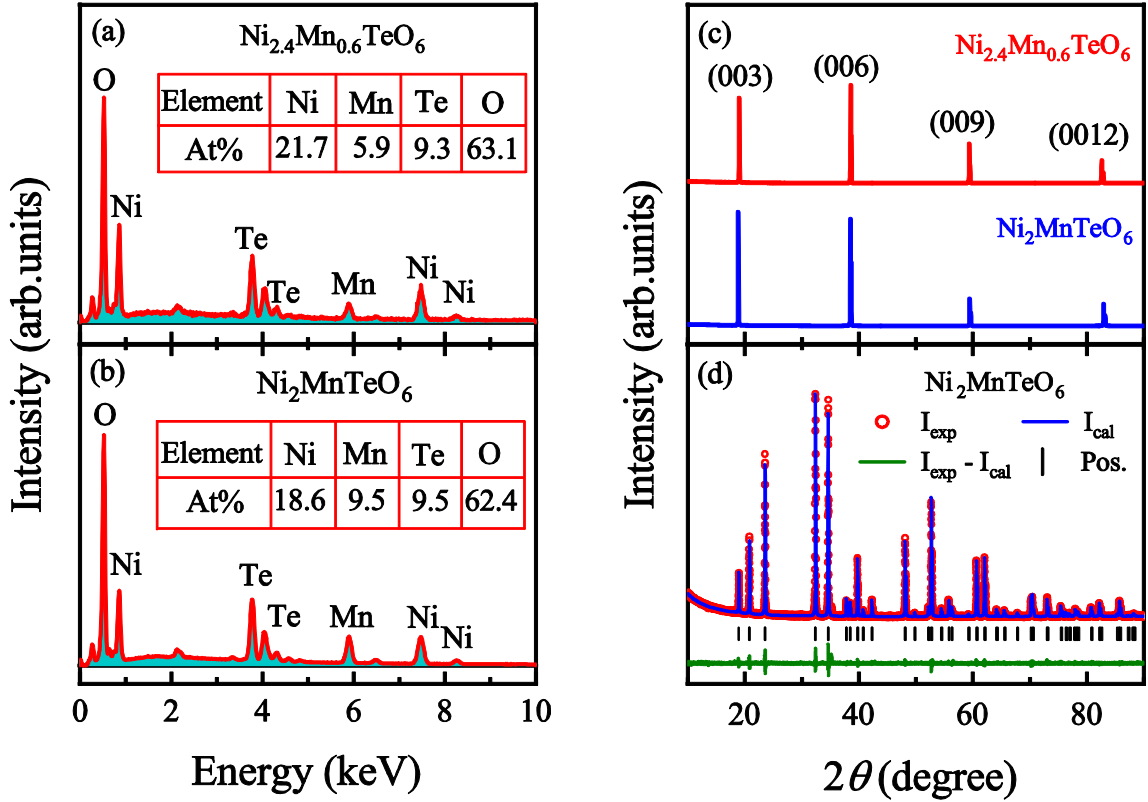

Fig. 2. The EDS spectrum of (a) $Ni_{2.4}Mn_{0.6}TeO_6$ and (b) $Ni_2MnTeO_6$ single crystals. The insets in (a)-(b) show the atomic ratio of Ni, Mn, Te, and O elements. (c) Room-temperature XRD patterns of $Ni_{2.4}Mn_{0.6}TeO_6$ and $Ni_2MnTeO_6$ single crystals. (d) Room-temperature XRD pattern and Rietveld refinement of crushed $Ni_2MnTeO_6$ single crystals. The observation ($I_{exp}$), calculation ($I_{cal}$), and their difference ($I_{exp} - I_{cal}$) are plotted in red open circles, blue and olive-green solid lines, respectively. The black vertical bars mark the Bragg position (Pos.) of $Ni_2MnTeO_6$.

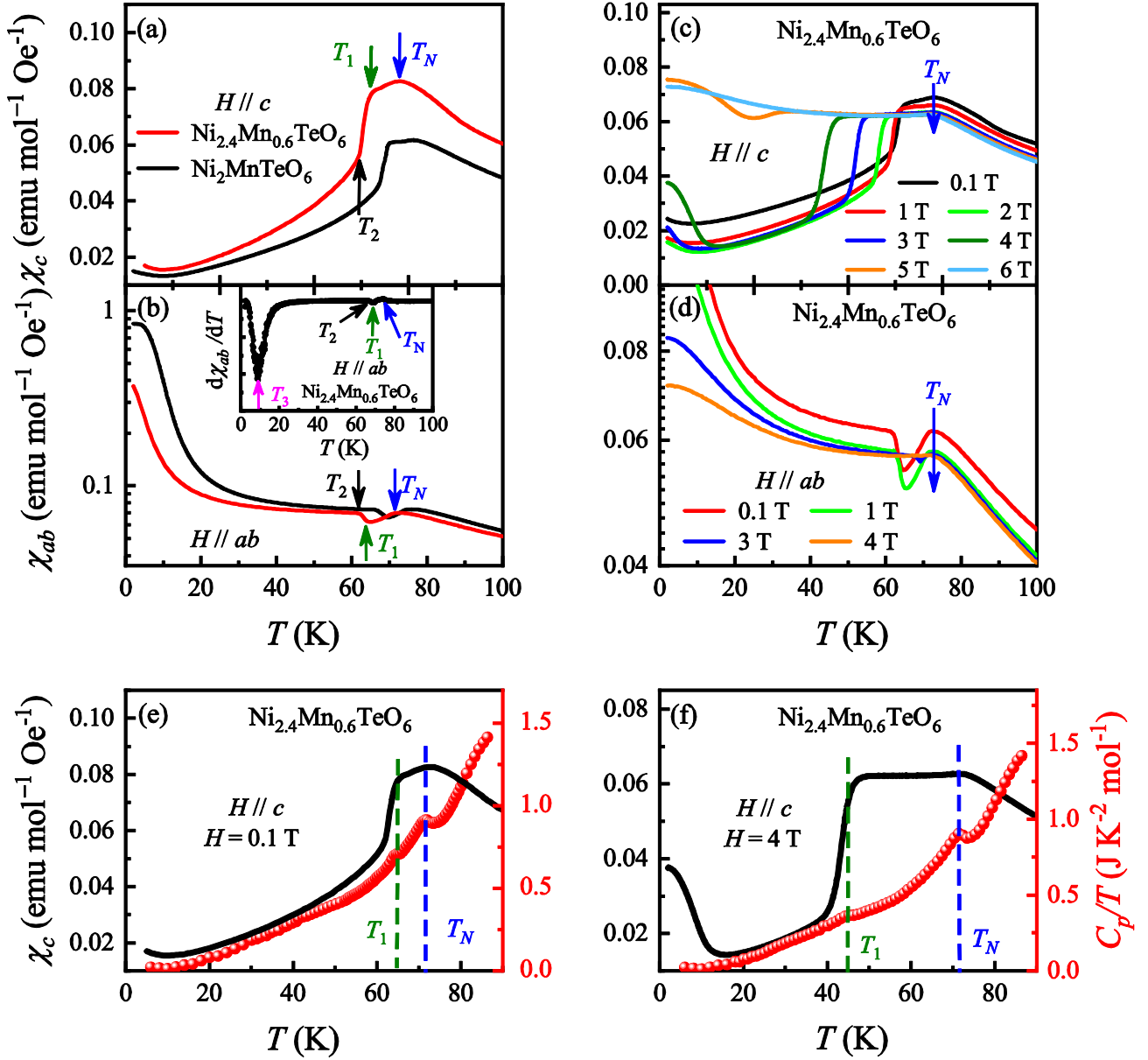

Fig. 3. The $T$-dependent magnetic susceptibility of $Ni_{2.4}Mn_{0.6}TeO_6$ and $Ni_2MnTeO_6$ for (a) $H // c$ and (b) $H // ab$ under zero-field-cooled mode with measuring field $H = 0.1$ T. The inset in (b) shows the $T$-derivative of magnetic susceptibility $d\chi_{ab}/dT$ of $Ni_{2.4}Mn_{0.6}TeO_6$ for $H // ab$. (c) and (d) are the $T$-dependent magnetic susceptibility of $Ni_{2.4}Mn_{0.6}TeO_6$ under different magnetic fields applied along and perpendicular to the $c$-axis, respectively. The $T$-dependent magnetic susceptibility and specific heat $C_p/T(T)$ measured in the magnetic field (e) $H = 0.1$ T and (f) $H = 4$ T. The blue arrows in (a)-(d) denote the Néel point $T_N$, and green and back arrows in (a)-(b) indicate anomalies in magnetic susceptibility. The blue and olive dashed lines in (e) and (f) indicate the magnetic transition points $T_N$ and $T_1$, respectively.

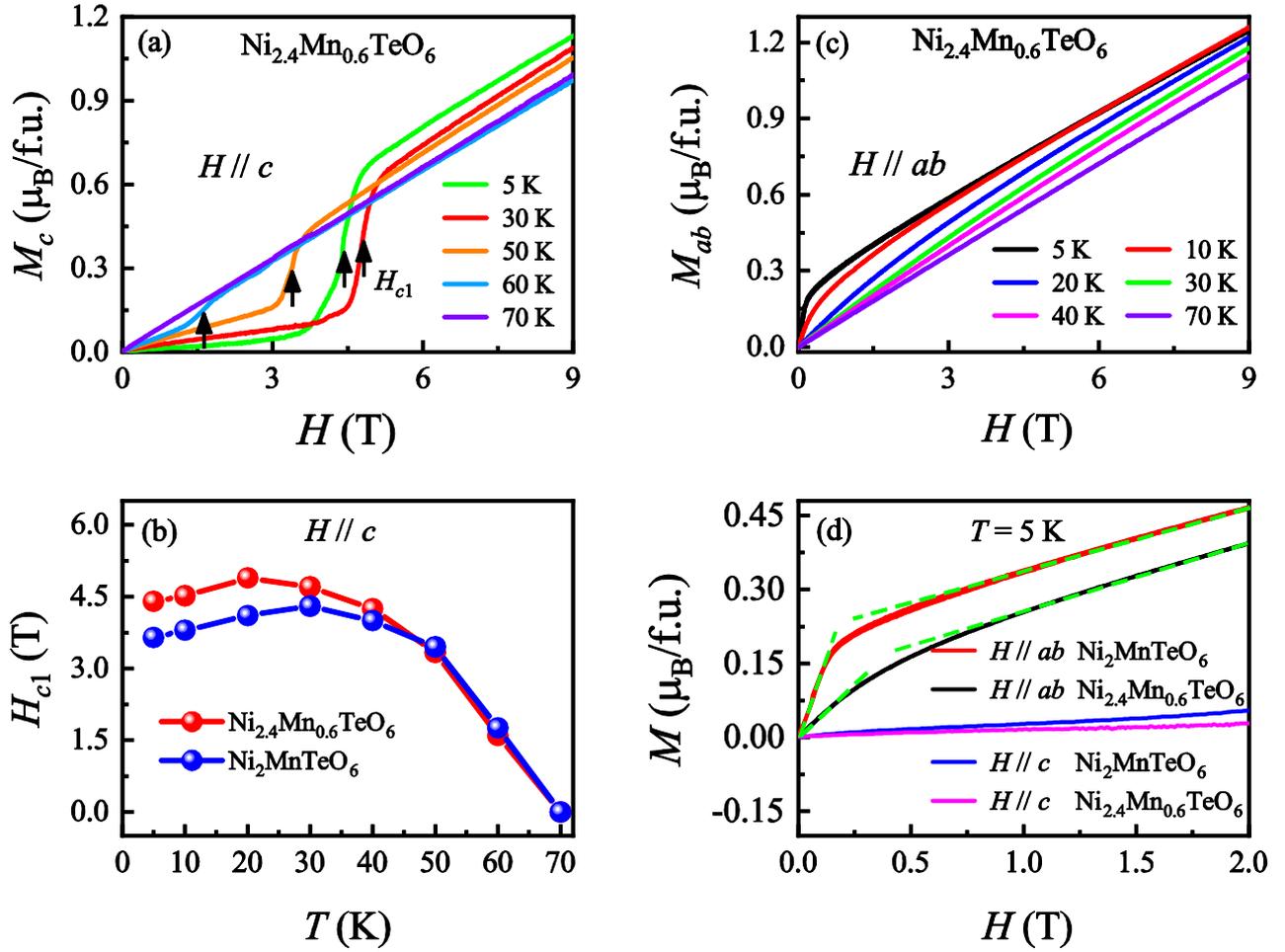

Fig. 4. The $H$ dependent magnetization $M$ of $Ni_{2.4}Mn_{0.6}TeO_6$ at selected $T$ under the $H$ up to 9 T for (a) $H // c$ and (c) $H // ab$. (b) The $T$ dependent critical field $H_{c1}$ of $Ni_{2.4}Mn_{0.6}TeO_6$ and $Ni_2MnTeO_6$ for $H // c$. (d) $H$ dependent magnetization of $Ni_{2.4}Mn_{0.6}TeO_6$ and $Ni_2MnTeO_6$ at $T = 5$ K under $H // c$ and $H // ab$. The black arrows in (a) indicate the critical magnetic field point of metamagnetic transition. The green dashed lines in (d) are used to show the linear relationship.

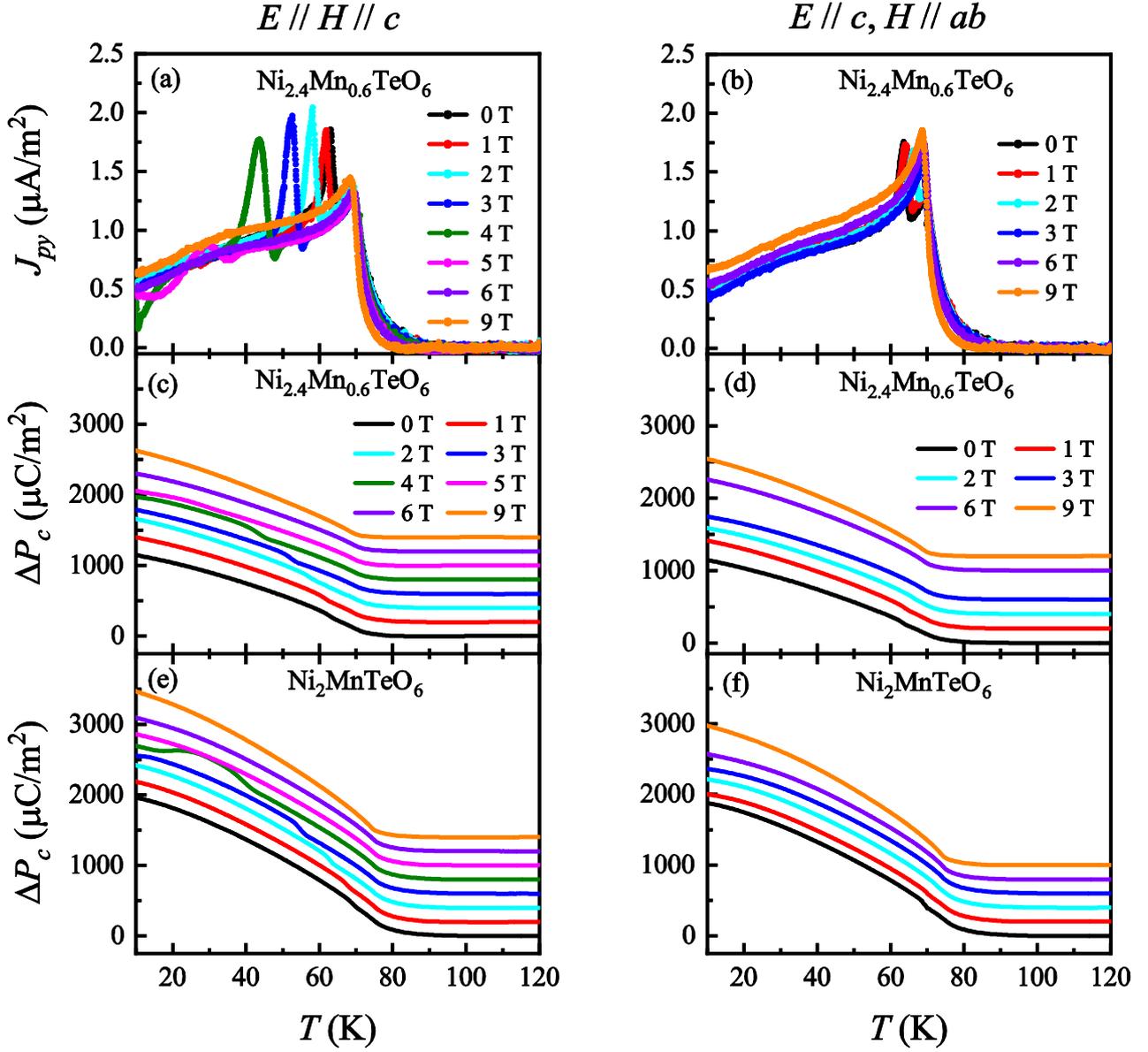

Fig. 5. The *T*-dependent pyroelectric current density ($J_{py}$) along the *c*-axis of Ni$_{2.4}$Mn$_{0.6}$TeO$_6$ for (a) *H* // *c* and (b) *H* // *ab*. (c)-(f) The *T* dependence of the change in electric polarization $\Delta P_c$ along the *c*-axis under (c) *H* // *c*, (d) *H* // *ab* for Ni$_{2.4}$Mn$_{0.6}$TeO$_6$ and (e) *H* // *c*, (f) *H* // *ab* for Ni$_2$MnTeO$_6$. The data in (c)-(f) are vertically offset for clarity.

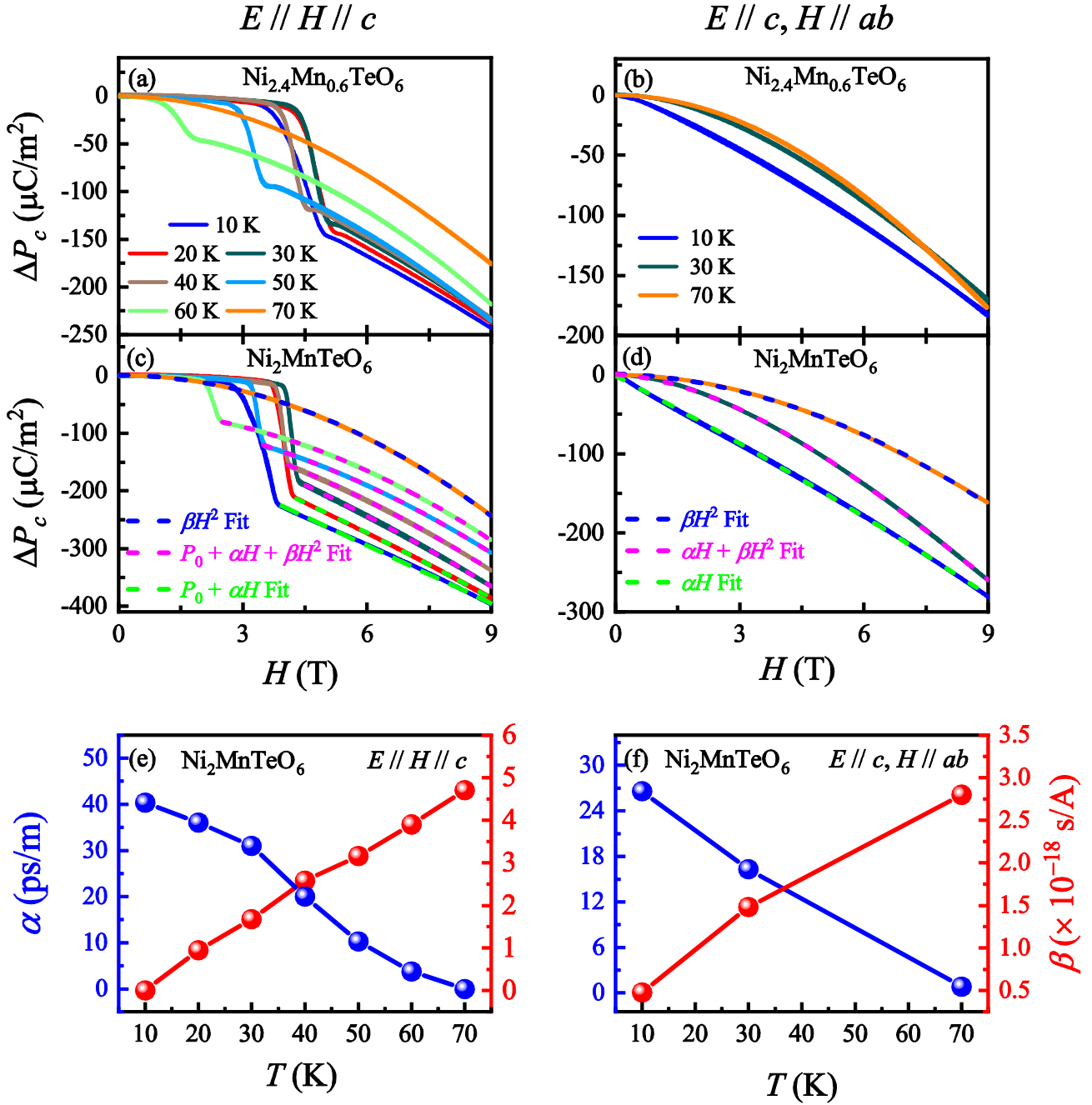

Fig. 6. The $H$ dependence of the change in electric polarization $\Delta P_c$ along the $c$-axis at various temperatures under (a) $H \mathbin{/\mkern-6mu/} c$, (b) $H \mathbin{/\mkern-6mu/} ab$ for $Ni_{2.4}Mn_{0.6}TeO_6$, and (c) $H \mathbin{/\mkern-6mu/} c$, (d) $H \mathbin{/\mkern-6mu/} ab$ for $Ni_2MnTeO_6$. The $T$-dependent magnetoelectric coefficients $\alpha$ and $\beta$ of $Ni_2MnTeO_6$ obtained by fitting $\Delta P_c(H)$ curves under (e) $H \mathbin{/\mkern-6mu/} c$ and (f) $H \mathbin{/\mkern-6mu/} ab$.

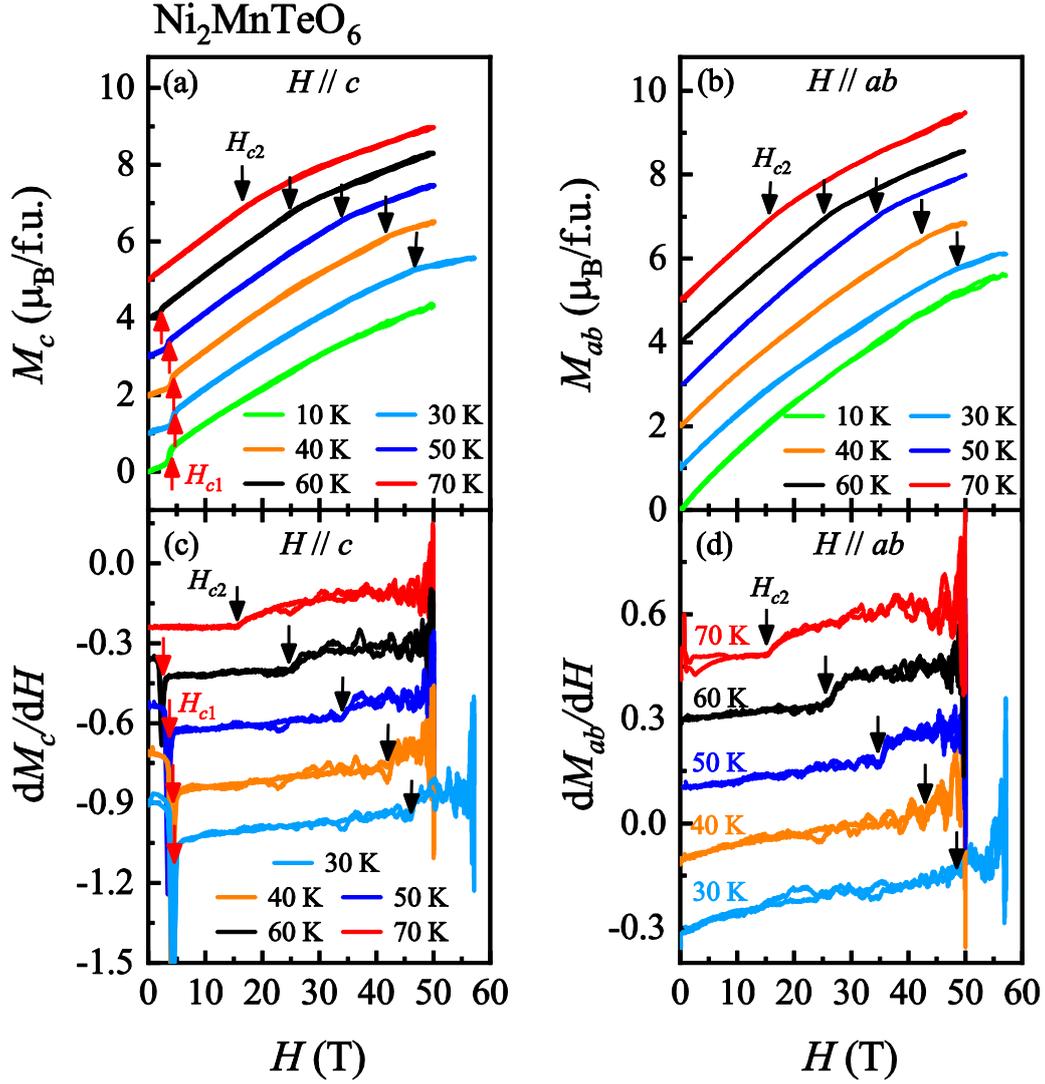

Fig. 7. The high-field magnetization of $Ni_2MnTeO_6$. The $H$-dependence of the magnetization (a) $M_c$ for $H // c$, (b) $M_{ab}$ for $H // ab$ at several selected $T$. The $H$-derivative of the magnetization (c) $dM_c/dH$ for $H // c$, and (d) $dM_{ab}/dH$ for $H // ab$. Curves in (a)-(d) are vertically shifted for clarity. The red and black arrows in (a), (c) and the black arrows in (b), (d) are used to indicate the magnetic transition points.

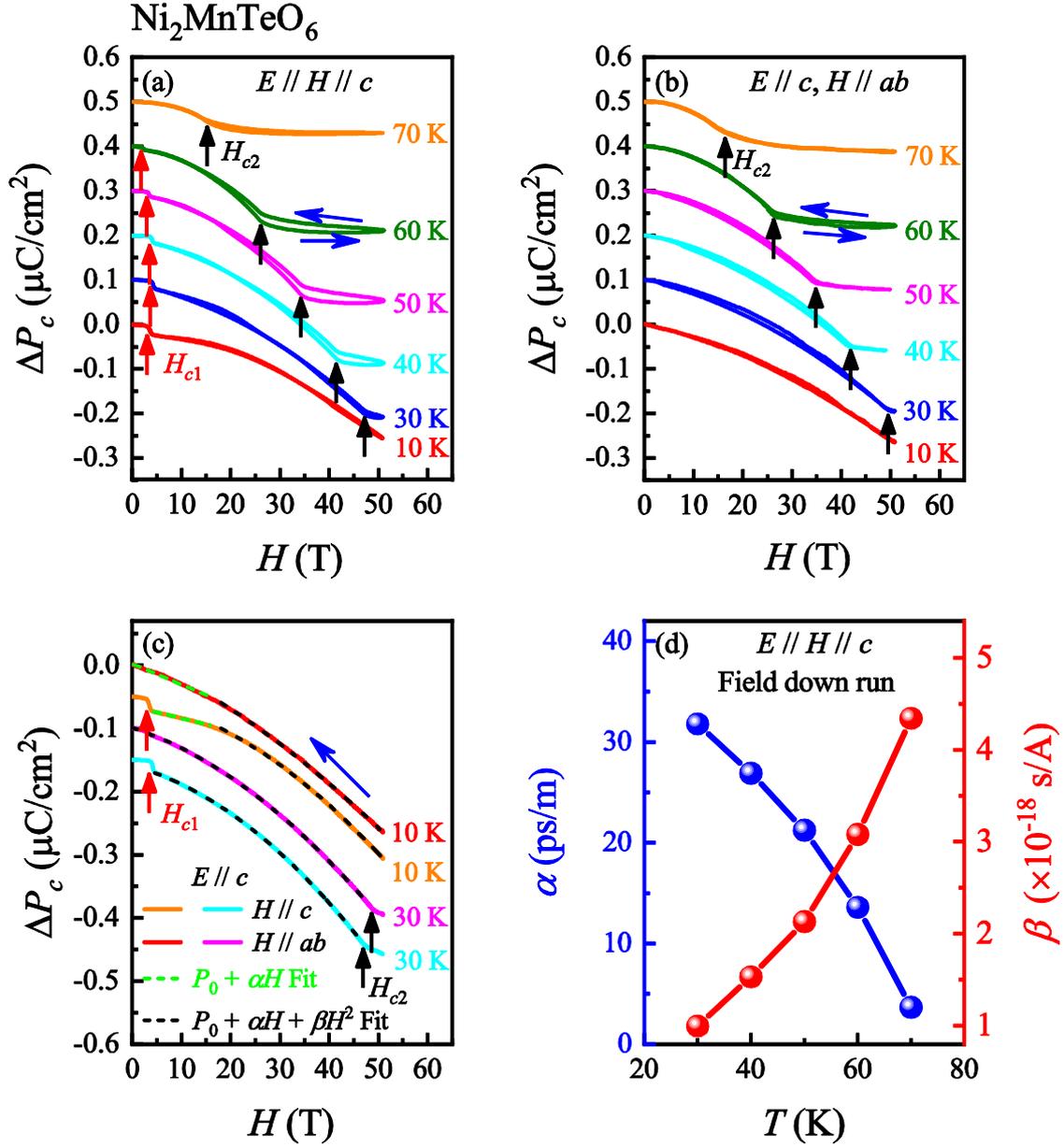

Fig. 8. The $H$-dependence of the change in electric polarization $\Delta P_c$ along the $c$-axis of Ni$_2$MnTeO$_6$ at various temperatures for (a) $H // c$, and (b) $H // ab$. (c) The $H$-dependent $\Delta P_c$ at selected temperatures with $H // c$ and $H // ab$ up to 52 T in the field-down run. (d) The $T$-dependent magnetoelectric coefficients $\alpha$ and $\beta$ obtained by fitting the $\Delta P_c(H)$ curves above 30 K the in field-down run. The dashed green and black lines in (c) are the fits to the $\Delta P_c$-$H$ curves with the function of $\Delta P \sim P_0 + \alpha H$ and $\Delta P \sim P_0 + \alpha H + \beta H^2$, respectively, where $P_0$, $\alpha$ and $\beta$ are the constants, noting that $P_0$ are zero for $H // ab$. Curves in (a)-(c) are vertically offset for clarity. The red and black arrows in (a),(c) and the black arrows in (b) are used to show the metamagnetic transition points. The blue arrows in (a)-(c) indicate the field-scan direction of the measurement.